\documentstyle{article}
\input epsf.sty

\begin{document}

\title{Variability model for the
            microquasar GRS 1915+105}

\author{ A. Janiuk$^{*}$, 
             B. Czerny\footnote{Nicolaus Copernicus Astronomical Center,
             Bartycka 18,
            00-716 Warsaw, Poland; agnes@camk.edu.pl; bcz@camk.edu.pl},
             A. Siemiginowska \footnote{ Harvard Smithsonian Center for Astrophysics, 60 Garden Street, MS-70, Cambridge MA 02138; aneta@head-cfa.harvard.edu}
and
             M.A. Abramowicz\footnote{ Department of Astronomy and Astrophysics, G\" oteborg University and Chalmers University of Technology, 
S-412 96 G\" oteborg, Sweden; marek@tfa.fy.chalmers.se }
\footnote{ Scuola Internazionale Superiore di Studi Avanzati, Via Beirut 2-4, I-34 014 Trieste, Italy}
}


\maketitle

\begin{abstract}
We analyse the radiation pressure instability of the standard, optically thisc disc and propose it as an explanation of the obseved variability of the superluminal microquasar GRS 1915+105. The quantitave agreement with the observed  outburst amplitudes is obtained under the assumption that substantial part of the energy is carried away by the jet.
\end{abstract}

\section{ Microquasar GRS 1915+105}

The X-ray source GRS 1915+105 discovered by GRANAT observatory
 exhibits both the superluminal jet (Mirabel 
\& Rodriguez 1994) and very complex variability pattern (Belloni et al. 1997).
The outburst X-ray luminosity is about $10^{39}$ erg/s, being of the order of 
Eddington luminosity in case of stellar mass black hole system.
The spectral analysis reveals the domination of disk-like component during the
outbursts, and therefore the variability of the source may be connected with 
disk evolution.

\section{ Instability in the accretion disc}

The local time behaviour of an accretion disk at a given radius is determined 
by the stability curve, i.e. the log $\dot m$ vs. log $\Sigma$ diagram
constructed for a stationary disk, where $\dot m$ is the accretion rate and
$\Sigma$ is the disk surface density. The parts of the curve with a negative 
slope describe the unstable branch and indicate the range of accretion rates
resulting in time dependent disk behaviour at this radius. The unstable part
 corresponds to the radiation pressure dominated region; including the
 advection term results in stabilizing the solution. 


\begin{figure}
\epsfxsize = 100 mm 
\epsfbox{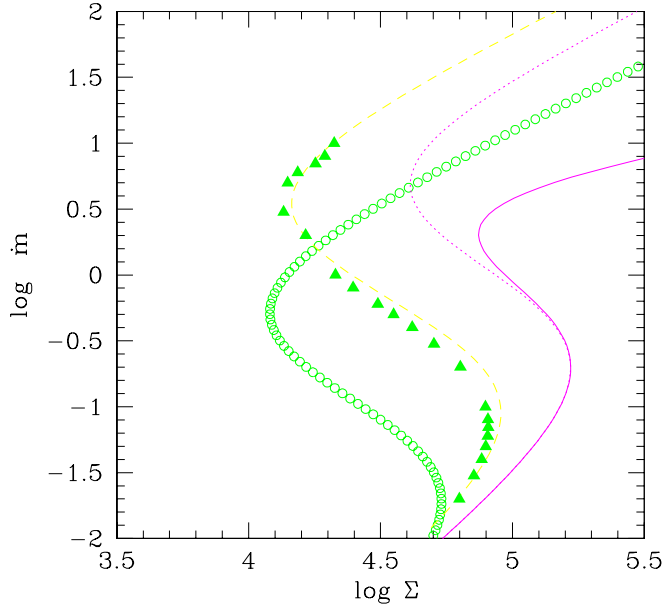}
\caption{
An example of the stability curves calculated at 10 $R_{Schw}$ for the mass
of the central object equal $10 M_{\odot}$ and  viscosity parameter $\alpha$ 
equal 0.01. Green triangles show the solution obtained with the model of full vertical structure of a disc and the circles mark the
standard analytic solution. The yellow line is obtained from the verticvally averaged solution with appropriate scaling coefficients. Red continous lines are the curves resulting from the vertically averaged equations: dotted line shows the effect of assumption that 50\% of the gravitational energy is released in the corona outside the disc and the  continuous line resulted from including  the jet.}
\end{figure}

The disc is unstable in the inner part. For the limited range of radii the 
limit cycle behaviour occurs and propagates outwards. The value of viscosity 
parameter does not change the outburst amplitude, but influences the duration 
of the limit cycle. 

\section{Jet ejection}

The observations of microquasar outbursts require that the whole limit
 cycle should run between $\dot m \sim 0.01$ and  $\dot m \sim 1.0$. In the model, the maximal disc luminosity is lowered under the assumption that jet ejected from the disc takes away a certain part of the total emitted power, depending on the $L/L_{Edd}$ ratio. We adopt a jet power parameterization similar to 
Nayakshin et al. (1999)

We assume that the heat generated within the disk at any radius 
is either stored 
temporarily within it or removed by: (i) radiation (ii) advection (iii) jet.
As the spectral observations show the presence of hard X-ray tail, we also 
assume the existence of hot corona, with moderate contribution to the energy dissipation ($f_{cor}=0.5$).

\section{Theoretical lightcurves}

We adopt the mass supply to the inner part of the disk as a 
model parameter and follow the time evolution of the disk under the
radiation pressure instability, as described in Siemiginowska et al. 1996.
The amplitude of the global outburst depends mostly on the shape
of the stability curve in the innermost part of the disk.
The duration of the limit cycle, on the other hand, 
is basically determined by 
the viscous timescale at the outer radius of the instability zone
in a stationary model, i.e. on the stationary accretion rate, radial
extension of the instability zone for that accretion rate and the
viscosity parameter $\alpha$.

 
\begin{figure}
\epsfxsize = 100 mm 
\epsfbox{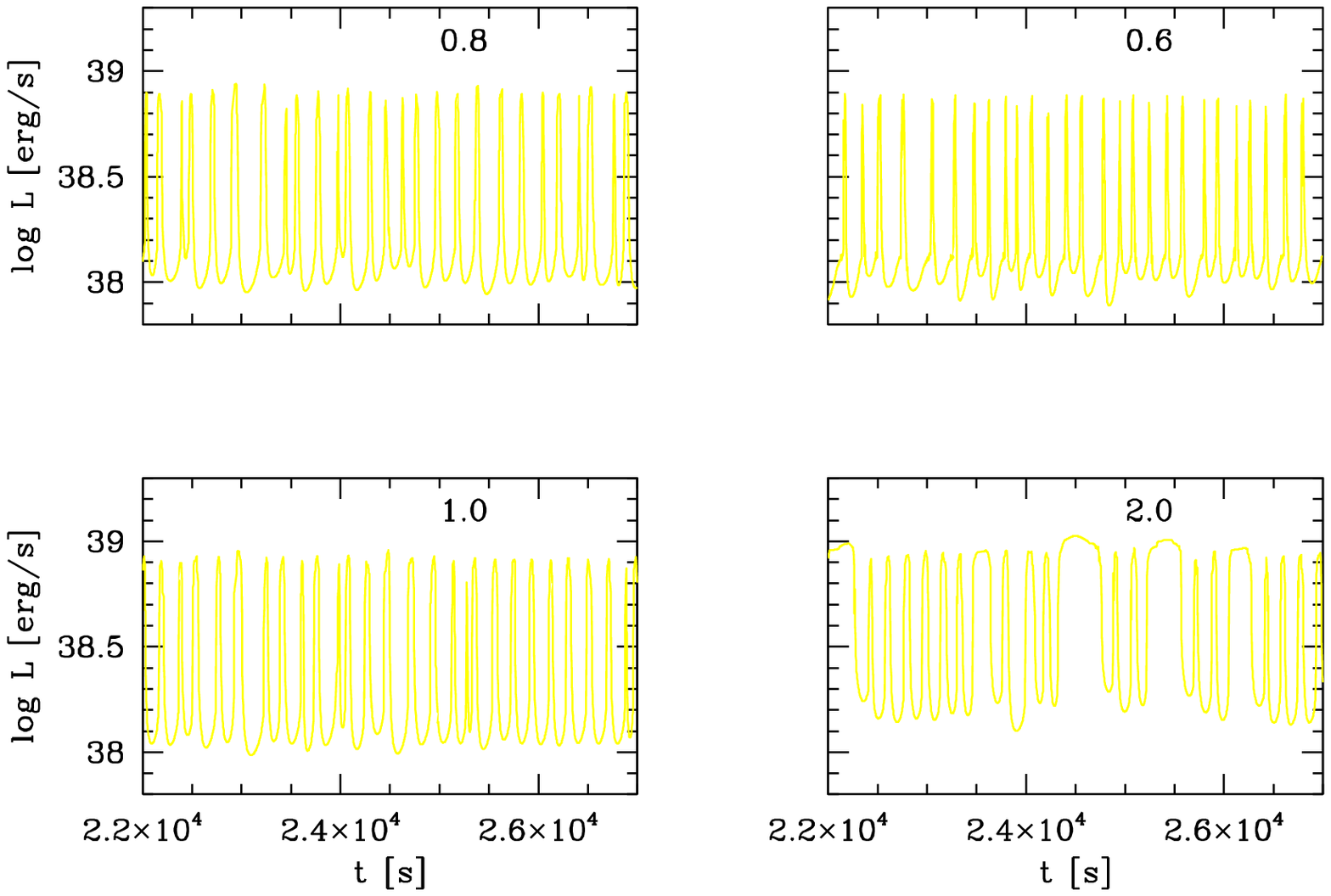}
\caption{
The lightcurves  calculated for $M=10 M_{\odot}$,
 $\alpha = 0.01$ and four
values of the external accretion rate: $0.6\times 10^{-7}$, $0.8\times 10^{-7}$, $1.0\times 10^{-7}$ and $2.0\times 10^{-7} [M_{\odot}/yr]$ .}
\end{figure}

\section{Summary}

 We used a physically viable
instability due to radiation pressure to model the time dependent behaviour of the 
optically thick disk itself. The presence of the hot optically thin medium was
accounted for by assuming assuming that a certain part of gravitational energy  ($f_{cor}=0.5$) is released in hot corona outside the disc.  
The mechanism operates for accretion rates above $\dot M_{0}/ \dot M_{Edd} \sim 0.16$ 
which is
in agreement with the observations of GRS 1915+105 - the source
does not exhibit the outburst if the mean luminosity temporarily 
drops below $2.1 \times 10^{38}$ erg/s. 
The viscosity coefficient of order of 0.01
is appropriate to model the typical outburst duration.
We conclude that the radiation pressure instability is a promising 
model of the basic instability mechanism underlying the observed
variability of this microquasar.

\vspace{1cm}

\section{References}

\vspace{0.5cm}
\noindent
Abramowicz M. A., Czerny B., Lasota J.-P., Szuszkiewicz E., 1988,
 ApJ, 332, 646\\
Belloni T., Mendez M., King A. R., van der Klis M., van Paradijs J., 1997, ApJ, 479, L145\\  
Mirabel I.F., Rodriguez L.., 1994, Nature, 371, 46\\
Nayakshin S., Rappaport S., Melia F., astro-ph/9905371\\
Siemiginowska A., Czerny B., Kostyunin V., 1996, ApJ, 458, 491\\
Taam R.E., Chen X., Swank J.H., 1997, ApJ, 485, L83\\

\end{document}